\pdfoutput=1
\documentclass[
  ,final            
  ]
  {aipproc}
\layoutstyle{8x11single}

\usepackage[cp1251,engukr]{statphys09}
\usepackage{amsmath}
\begin{document}

\title{Network harness: bundles of routes in public transport networks}


\author{B. Berche}{
  address={Statistical Physics Group, Institut Jean Lamour, UMR CNRS 7198, Nancy Universit\'{e}, 54506 Vand\oe uvre les Nancy Cedex, France} }

\author{C.~von Ferber}{
  address={Coventry University, Applied Mathematics Research Centre,
CV1 5FB Coventry, UK},
 altaddress={Universit\"{a}t Freiburg, Physikalisches Institut, D-79104 Freiburg, Germany}}

\author{T.~Holovatch\footnote{holtaras@lpm.u-nancy.fr}\hspace{0.5em}}{
altaddress={Coventry University, Applied Mathematics Research
Centre, CV1 5FB Coventry, UK},
  address={Statistical Physics Group, Institut Jean Lamour, UMR CNRS 7198, Nancy Universit\'{e}, 54506 Vand\oe uvre les Nancy Cedex, France} }

\begin{abstract}
 Public transport routes sharing the same grid of streets and tracks
are often found to proceed in parallel along shorter or longer
sequences of stations. Similar phenomena are observed in other
networks built with space consuming links such as cables, vessels,
pipes, neurons, etc. In the case of public transport networks (PTNs)
this behavior may be easily worked out on the basis of sequences of
stations serviced by each route. To quantify this behavior we use
the recently introduced notion of network harness. It is described
by the harness distribution $P(r,s)$: the number of sequences of $s$
consecutive stations that are serviced by $r$ parallel routes. For
certain PTNs that we have analyzed we observe that the harness
distribution may be described by power laws. These power laws
observed indicate a certain level of organization and planning which
may be driven by the need to minimize the costs of infrastructure
and secondly by the fact that points of interest tend to be
clustered in certain locations of a city. This effect may be seen as
a result of the strong interdependence of the evolutions of both the
city and its PTN.

To further investigate the significance of the empirical results we
have studied one- and two-dimensional models of randomly placed
routes modeled by different types of walks. While in one dimension
an analytic treatment was successful, the two dimensional case was
studied by simulations showing that the empirical results for real
PTNs deviate significantly from those expected for randomly placed
routes.
\end{abstract}

\maketitle


\section{Introduction}

A variety of different phenomena have in recent years been analyzed
in the context of complex network theory [1, 2]. Usually, the focus
is on the network topology while the study of specific features as
e.g. the network load or real-space correlations is mostly left
aside.\footnote{Paper presented at the Conference ``Statistical
Physics: Modern Trends and Applications'' (23-25 June 2009, Lviv,
Ukaine) dedicated to the 100th anniversary of Mykola Bogolyubov
(1909-1992).}

Analyzing statistical properties of public transport networks (PTNs)
[3, 4, 5] a so-called harness effect which results from such spatial
correlations has recently been proposed [5, 6]. The latter may be
observed for networks on which a set of walks or paths is defined.
Indeed, public transport routes sharing the same grid of streets and
tracks define such a set of walks on this grid. Often these are
found to proceed in parallel along shorter or longer sequences of
stations. Similar phenomena are observed in other networks built
with space consuming links such as cables [7], vessels [8], pipes
[9], neurons [10], etc. In the case of PTNs, a quantitative
description of sequences of stations that are served by several
routes may be performed in the form of the harness distribution
$P(r,s)$: the number of maximal found on the network sequences of
$s$ consecutive stations that are serviced by $r$ parallel routes.
In a recent empirical analysis of the harness effect on PTNs of 14
major cities in the world [5, 6] it was found that for certain
cities the behavior of the function $P(r,s)$ is described by power
laws. These observed power laws indicate a certain level of
organization and planning which may be driven by the need to
minimize the costs of infrastructure and secondly by the fact that
points of interest tend to be clustered in certain locations of a
city.

In the present paper, to further investigate the significance of the
empirical results we have studied one- and two-dimensional models of
randomly placed routes modeled by different types of walks. The
setup of the paper is the following. First we recall some results of
the empirical analysis. We then proceed solving a simple 1d model of
a growing non-correlated network that allows an analytical
treatment. Finally we analyze different types of growing networks on
the 2d square lattice and perform numerical simulations and measure
the resulting harness effect as function of the model parameters.
Conclusions are given in the last section.

\section{Results of an empirical analysis}

In a previous study we have analyzed the PTNs of 14 major cities
that have different geographical, historical and cultural background
[5, 6]. For some of these we observed non-vanishing harness
distributions $P(r,s)>0$ even for long sequences $s$ and high
numbers of routes $r$. This is what we call a "strong" harness
effect (examples are Sao Paolo, Hong Kong, Istanbul, Los Angeles,
Rome, Sydney, Taipei, Moscow, London; some are shown in Fig. 1). For
other PTNs the maximal values of $s$ and $r$ with $P(r,s)>0$ were
found to be smaller than 10 (Berlin, Paris, Dallas, Duesseldorf,
Hamburg) - this we call a "weak" harness effect (Fig. 2). It is
important to note that the division into these two classes does not
correlate with either the average number of routes $R$ in the PTN or
their average length $S$.
\begin{figure}[*h]
\begin{tabular}{cc}
\includegraphics[width=.47\textwidth]{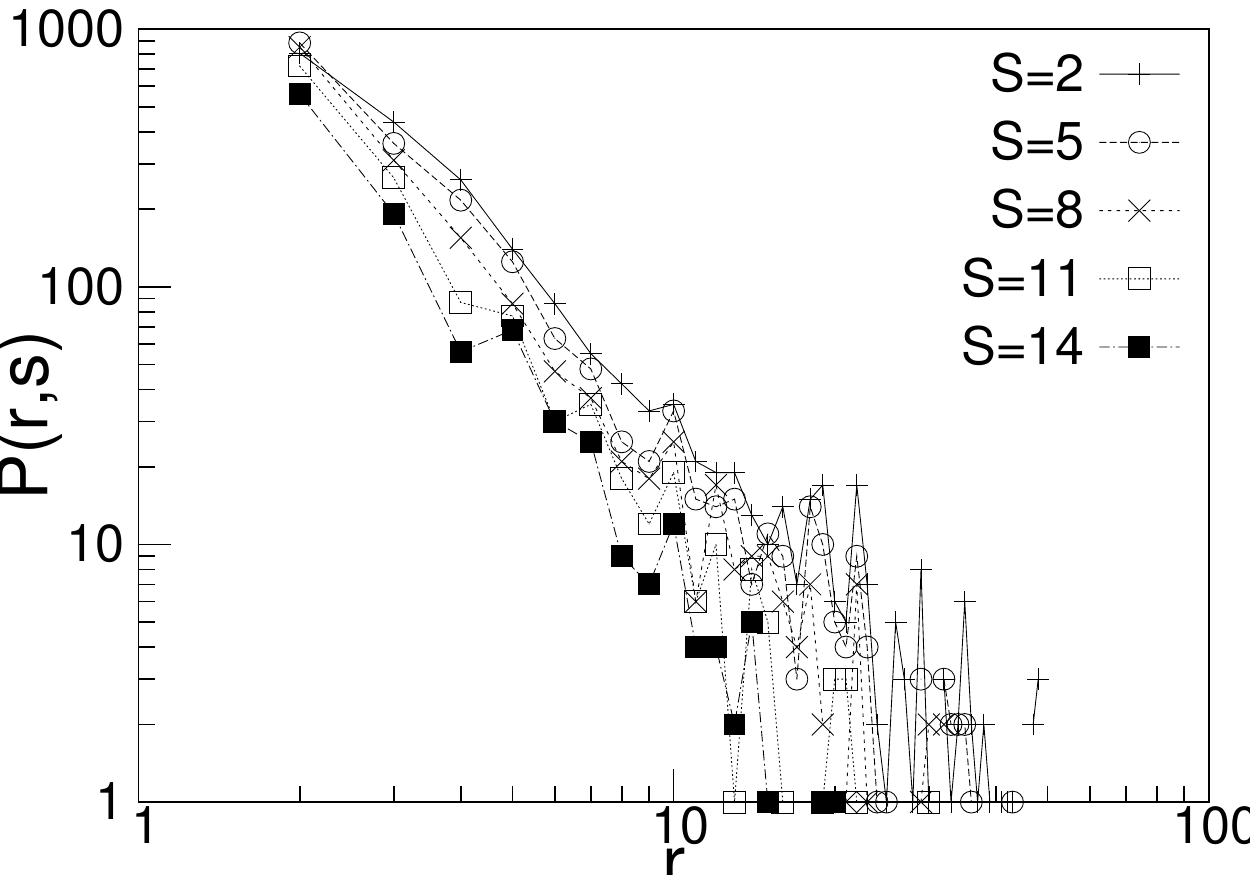} &
  \includegraphics[width=.47\textwidth]{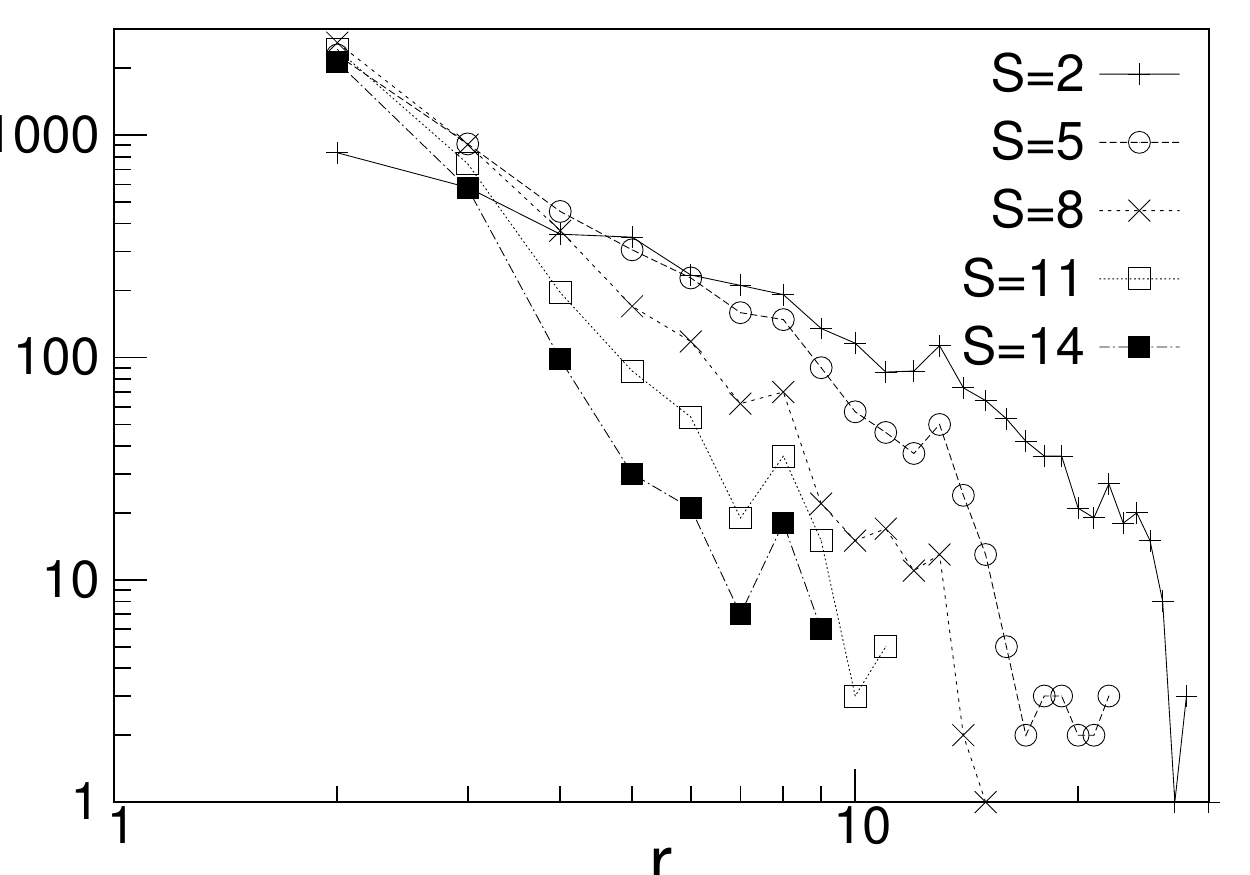}\\
  (a) & (b)\\
  \includegraphics[width=.47\textwidth]{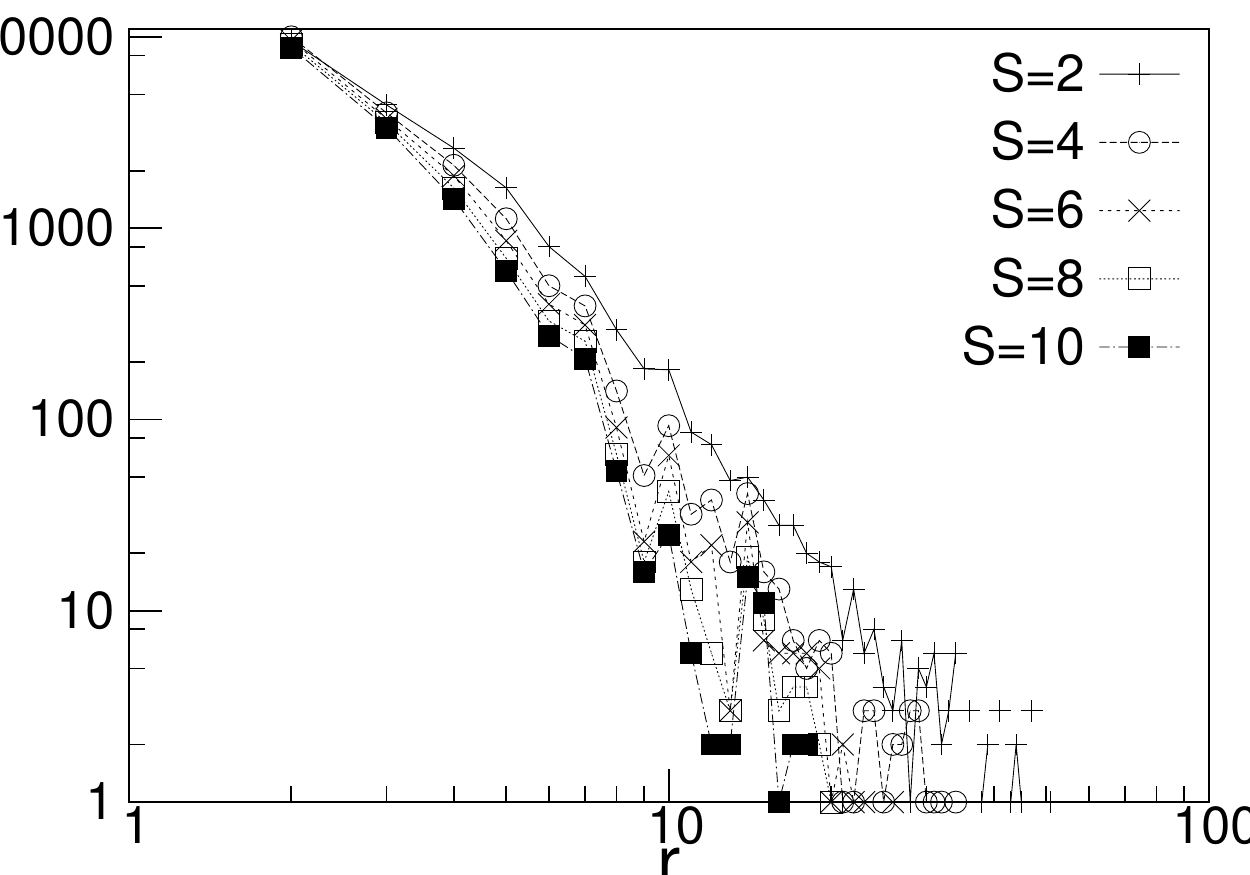} &
  \includegraphics[width=.47\textwidth]{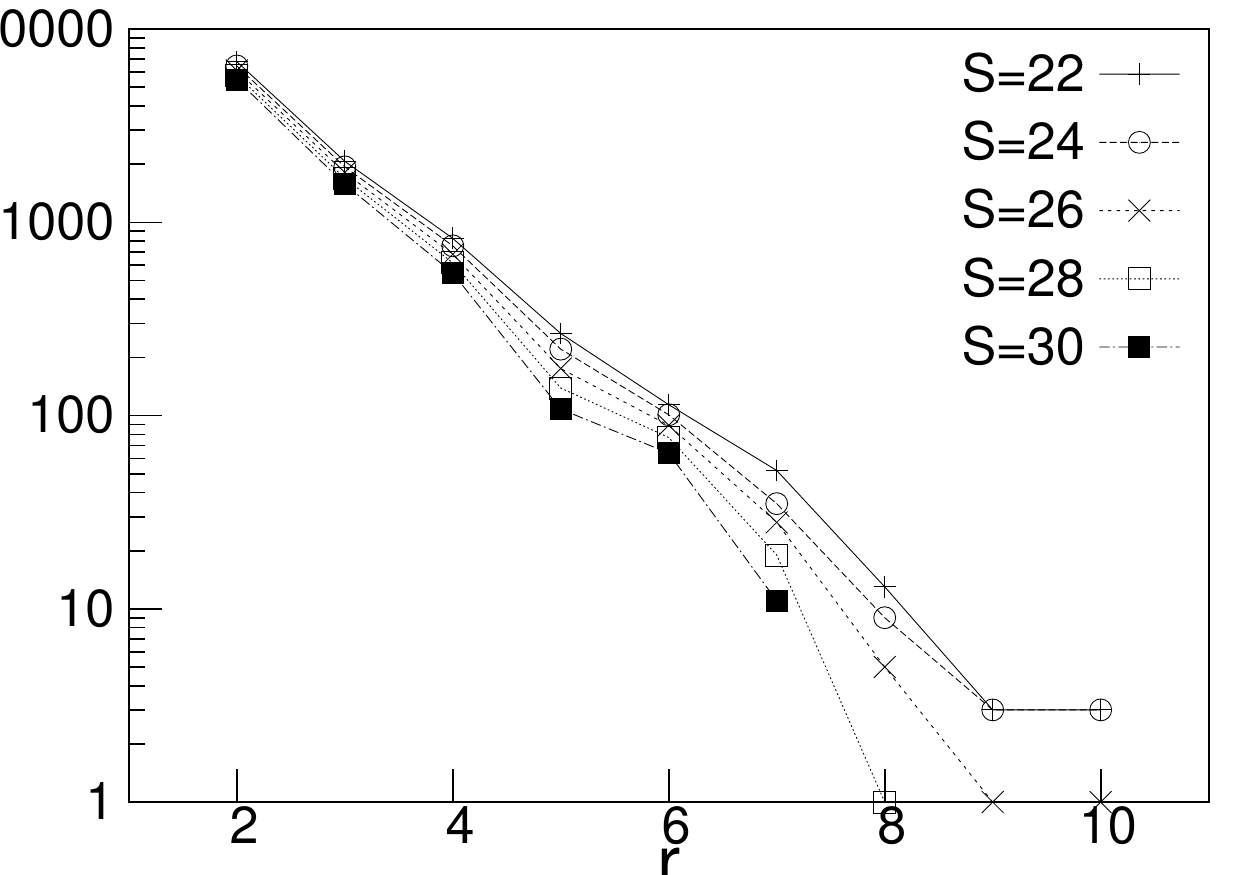}\\
  (c) & (d)
  \end{tabular}
  \caption{$s$-cumulative harness distribution $P_c(r,\hat s)$ as
  function of $r$ for fixed $\hat s$. Log-log for Istanbul (a) and Taipei (b) $\hat s
  = 2,5,8,11,14$. Log-log for Los Angeles $\hat s = 2,4,6,8,10$ (c) and
  log-lin for Los Angeles $\hat s=22,24,26,28,30$ (d).}
\end{figure}
 Another result is that similar to the
node-degree distributions
 [5, 6] we observe that the harness
distribution $P(r,s)$ for some of the cities (Sao Paolo, Hong-Kong,
Istanbul, Los Angeles, Rome, Sydney) may be described by a power
law:
 \begin{equation}
  \label{f1}
  P(r,s)\sim r^{-\gamma _{s}}\textrm{, \hspace{1em} for \hspace{0.2em} fixed \hspace{0.2em}
  s,}
  \end{equation}
whereas the PTNs of other cities (Taipei, Moscow, London, Hamburg)
are better described by an exponential decay:
 \begin{equation}
  \label{f2}
  P(r,s)\sim e^{-r/\hat r _{s}}\textrm{, \hspace{1em} for \hspace{0.2em} fixed \hspace{0.2em}
  s.}
  \end{equation}

We illustrate this behaviour in Figs. 1a,b showing the
  harness distribution for Istanbul
  and for Taipei. In some cases (e.g. for Rome and Los Angeles)
there is a crossover between regimes (1) and (2) at larger $s$ as
shown this for the PTN
  of Los Angeles (Fig.1c). Here, one can see that for small values
  of $s$ the results are better described by a power law (1).
  With increasing $s$ a tendency to an exponential decay (2) appears
  (Fig. 1d). This is less obvious for other cities analyzed, however in
  all cases the harness distribution
  $P(r,s)$ as function of $r$ decays faster for longer sequence lengths
and while also attaining a more pronounced curvature.

Note that in Fig. 1 we plot $s$-cumulative distributions $P_c(r,\hat
s)$ where a sequence with maximal length $s=9$ will be counted once
as a sequence of length $\hat s=9$ and twice as a sequence of length
$\hat s = 8$ etc:
\begin{equation}\label{f2b}
P_c(r,\hat s) = \sum_{s=\hat s}^{S}(s+1-\hat s)P(r,s)
\end{equation}
It may be surprising that these curves e.g. for Taipei (Fig. 1b)
intersect for low values of $r$. We will discuss this effect below.

  For PTNs for which the harness distribution follows a power law (1)
the corresponding exponents $\gamma _{s}$ are found in the range of
$\gamma _{s} = 2 \div 4$. For those distributions with an
exponential decay the scale $\hat r _{s}$ (see eq.(2) varies in a
range $\hat r _{s} = 1.5 \div 4$. The power laws observed for the
behavior of $P(r,s)$ indicate a certain level of organization and
planning which may be driven by the need to minimize the costs of
infrastructure and secondly by the fact that points of interest tend
to be clustered in certain locations of a city. Note that this
effect may be seen as a result of the strong interdependence of the
evolutions of both the city and its PTN. We want to emphasize that
the harness effect is a feature of the network given in terms of its
routes but it is invisible in any of the complex network
representations of public transport networks presented so far, such
as L-space [3], P-space [11] or B-space [5, 12]. It is possible,
that the notion of harness may be useful also for the description of
other networks with similar properties. On the one hand, the harness
distribution is closely related to distributions of flow and load on
the network. On the other hand, in the situation of space-consuming
links (such as tracks, cables, neurons, pipes, vessels) the
information about the harness behavior may be important with respect
to the spatial optimization of networks. A generalization may be
readily formulated to account for real-world networks in which links
(such as cables) are organized in parallel over a certain spatial
distance. While for the PTN this distance is simply measured by the
length of a sequence of stations, a more general measure would be
the length of the contour along which these links proceed in
parallel.

For the cities observed no correlation appears to occur between the
harness distribution behavior and other well-known network
characteristics that were analyzed, as for example the node-degree
distribution of PTNs [5].

\begin{figure}[h]
  \includegraphics[width=.47\textwidth]{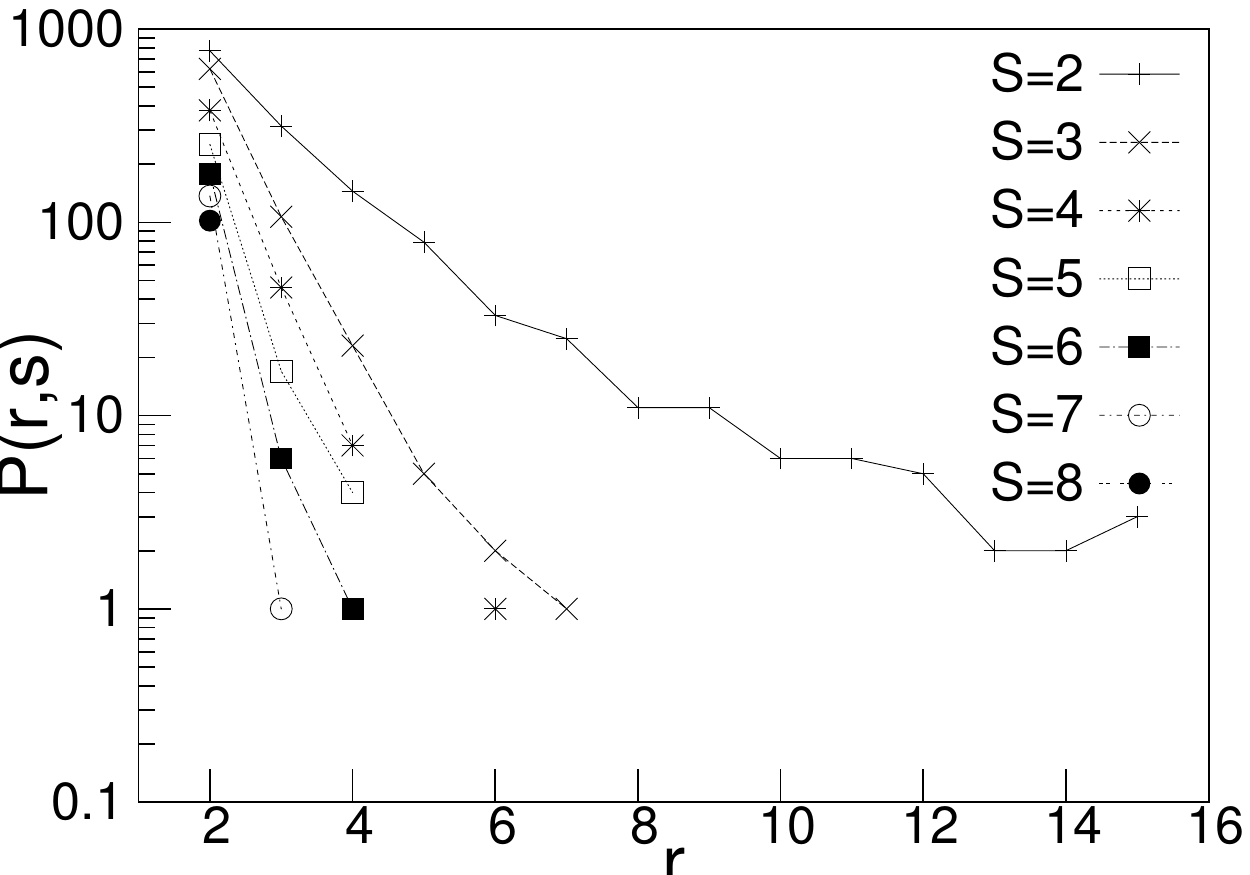}
  \caption{$s$-cumulative harness distribution $P_c(r,\hat s)$ as
  function of $r$ for fixed $\hat s$ for Paris ($\hat s = 2\div 8$).
$P_c(r,\hat s)=0$ if $\hat s>2$ for $r>7$ and if $\hat s>7$ even for $r>2$.}
\end{figure}

However, the extent to which harness properties are expressed may
obviously play a role for the attack vulnerability of a PTN.
Interestingly, our previous investigations have shown that the Paris
PTN is most resilient to any type of random or directed attacks (in
terms of percolation concepts) among all analyzed PTNs [13]. At the
same time it exhibits the "weakest" behavior with respect to the
harness effect (Fig. 2). One may expect such a result: routes that
do not share the same streets are more resilient. However, for other
cities no apparent correlation between harness effect behavior and
their vulnerability has been found so far.

As noted above the interesting question to answer is: are there
any structural evolution purposes behind this effect, or can it be
found just as well within simple random scenarios.

\section{Analytic results and modeling in 1d}

Let us first investigate a network model with routes placed randomly in one dimensional
space. Although being very simple this model can mimic a harness
effect and as we will see below, it allows an analytical solution. The
model is formulated in the following way:
\newline

\begin{figure}[h]
  \includegraphics[height=.05\textheight]{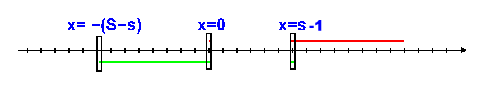}
  \caption{$R=2$ routes given as simple sequences of $S=15$ consecutive sites
 are placed at random on a line with $N$ sites.  }
\end{figure}

 The left terminals of $R$ routes of length $S$ are placed at random on
  a line with N sites, with periodic boundary conditions. E.g.
in Fig.3 we show two routes of length $S=15$ with left terminals at $x=0$ and $x=-8$.
We define the route density as $\rho = R/N$ routes per site.

  The distribution of left-terminals on a given site, e.g. site $x$
  will be $P_{x}(r)$: the probability that $r$ routes have their
  left-terminal on site $x$
  \begin{equation}
  \label{f3}P_{x}(r)= \begin{pmatrix}
  R\\
  r\\
\end{pmatrix} \cdot (\frac{1}{N})^{r} \cdot (1-\frac{1}{N})^{R-r},
  \end{equation}
  where the first term counts the number of ways to select $r$ of $R$ routes, the second term is
  the probability that the $r$ left terminals lie on site $x$ and the third
  term is the probability that no left terminal of an unselected route lies
on site $x$. In other words, by definition $P_{x}(r)$ is a binomial
distribution. For $ N \rightarrow \infty $, but fixed $\rho = R/N$
this
  distribution has the limiting behavior of a Poisson distribution:
  \begin{equation}
  \label{f4}
  P_{x}(r) \approx e^{-\rho}\cdot \frac{\rho^{r}}{r!}.
  \end{equation}

  Let us now calculate the probability that there is a sequence of
  maximal length $s$ and maximal width $r$ of $r$ routes in parallel
  between sites $x=0$ and $x=s$. This implies that at least one of
  the $r$ routes starts at $x=0$ and at least one of the routes ends
  at $x=s-1$. The latter route then
  starts at $x=-(S-s)\equiv -\bar s$ (Fig. 3). The other $r-2$
  routes may start anywhere in between $-\bar s\leq x \leq 0$. We
  denote the number of routes starting at $x=0$ as $r_{0}$, those
  starting at $x<0$ as $r_{-x}$.

In the limit $R \gg r$ and $N \gg \bar s$ we may consider
$P_{0}(r_{0})$, $P_{1}(r_{1})$, ... ,$P_{\bar s}(r_{\bar s})$ as
independent probabilities (this is not true for small systems,
however, as we see later the correlation between these probabilities
is negligible for the cases studied here. The overall probability to
find a sequence of length $s$ and width $r$ starting at $x=0$ is
then the sum over all combinations leading to the result:
  \begin{eqnarray}
  \nonumber
   P_{0}(r,s) &=& \sum^{(r)}_{\{r_i\}, r_{0}\geq 1,\hspace{0.1em} r_{\bar s} \geq 1}
P(r_{0})\cdot P(r_{1})\cdot... \cdot P(r_{\bar s}) =
\sum^{(r)}_{\{r_i\},\hspace{0.1em} r_{0}\geq 1,\hspace{0.1em} r_{\bar s} \geq1}
e^{-\bar s \rho}\cdot
\frac{\rho^{r_{0}}}{r_{0}!}\cdot\frac{\rho^{r_{1}}}{r_{1}!}\cdot ...
\cdot\frac{\rho^{r_{\bar s}}}{r_{\bar s}!} =
\\ \label{f5}
&=& e^{-\bar s \rho}\cdot \rho^{r} \cdot
\sum^{(r)}_{\{r_i\},\hspace{0.1em} r_{0}\geq 1,\hspace{0.1em} r_{\bar s} \geq 1}
\frac{1}{r_{0}!\cdot r_{1}!\cdot... \cdot r_{\bar s}!},
\end{eqnarray}
where $ \sum^{(r)}_{\{r_i\}} $ denotes a sum over $\{r_i\}$ with $r = r_{0}+r_{1}+...+r_{\bar s}$.

Now, without the conditions $r_{0} \geq 1$ and $r_{\bar s} \geq 1$
this sum can be derived from:
  \begin{equation}
  \label{f6}
 (\bar s +1)^{r} =
(1+1+1+...+1)^{r} = \sum^{(r)}_{\{r_i\}}\frac{r!}{r_{0}!\cdot r_{1}!\cdot...
\cdot r_{\bar s}!}.
  \end{equation}

  The sum with these conditions however can be written as:
  \begin{equation}
  \label{f7}
 \sum^{(r)}_{\{r_i\},\hspace{0.1em} r_{0}\geq 1,\hspace{0.1em} r_{\bar s} \geq1}
\frac{1}{r_{0}!\cdot r_{1}!\cdot... \cdot r_{\bar s}!} =
\sum^{(r)}_{\{r_i\}}\frac{1}{r_{0}!\cdot r_{1}!\cdot... \cdot r_{\bar s}!} -
\sum^{(r)}_{\{r_i\},\hspace{0.1em} r_{0}=0}\frac{1}{r_{0}!\cdot
r_{1}!\cdot... \cdot r_{\bar s}!} -
  \end{equation}
$$ - \sum^{(r)}_{\{r_i\},\hspace{0.1em} r_{\bar s}=0}
\frac{1}{r_{0}!\cdot r_{1}!\cdot... \cdot r_{\bar s}!} +
\sum^{(r)}_{\{r_i\},\hspace{0.1em} r_{0}=0,\hspace{0.1em} r_{\bar s}=0}
\frac{1}{r_{0}!\cdot r_{1}!\cdot... \cdot r_{\bar s}!} =
\frac{(\bar s +1)^{r}}{r!} - 2\frac{(\bar s)^{r}}{r!} + \frac{(\bar
s)^{r}}{r!}.$$ Thus:
\begin{equation}
  \label{f8}
  P_{0}(r,s) =  e^{(-\bar
s\cdot\rho)}\cdot\rho^{r}\cdot[(\bar s+1)^{r} - 2(\bar s)^{r} +
(\bar s -1)^{r}]/r!.
  \end{equation}

With this formula we count sequences that start at $x=0$. To
receive the overall probability this is to be multiplied by $N$.
\begin{equation}
  \label{f9}
  P(r,s) = N\cdot e^{(-\bar
s\cdot\rho)}\cdot\rho^{r}\cdot[(\bar s+1)^{r} - 2(\bar s)^{r} +
(\bar s -1)^{r}]/r!.
  \end{equation}

  Simple arithmetic allows us to calculate the $s$-cumulative
  distributions $P_c(r,\hat s)$, see eq. (3).

As mentioned above, the probabilities we use are appropriate for
infinite systems. To test their validity for finite cases we
performed some simple simulations.
\begin{figure}[t]
\begin{tabular}{cc}
  \includegraphics[width=.47\textwidth]{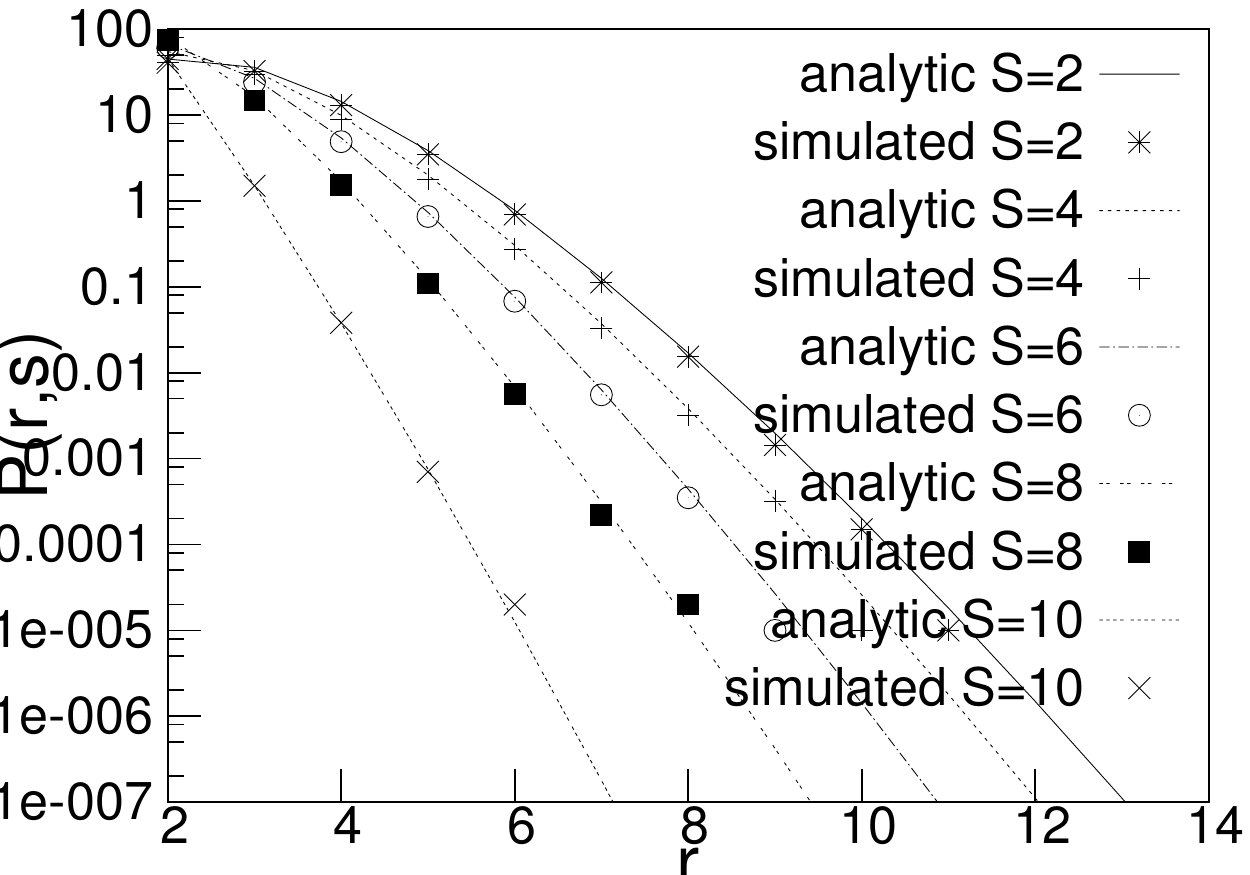} &
  \includegraphics[width=.47\textwidth]{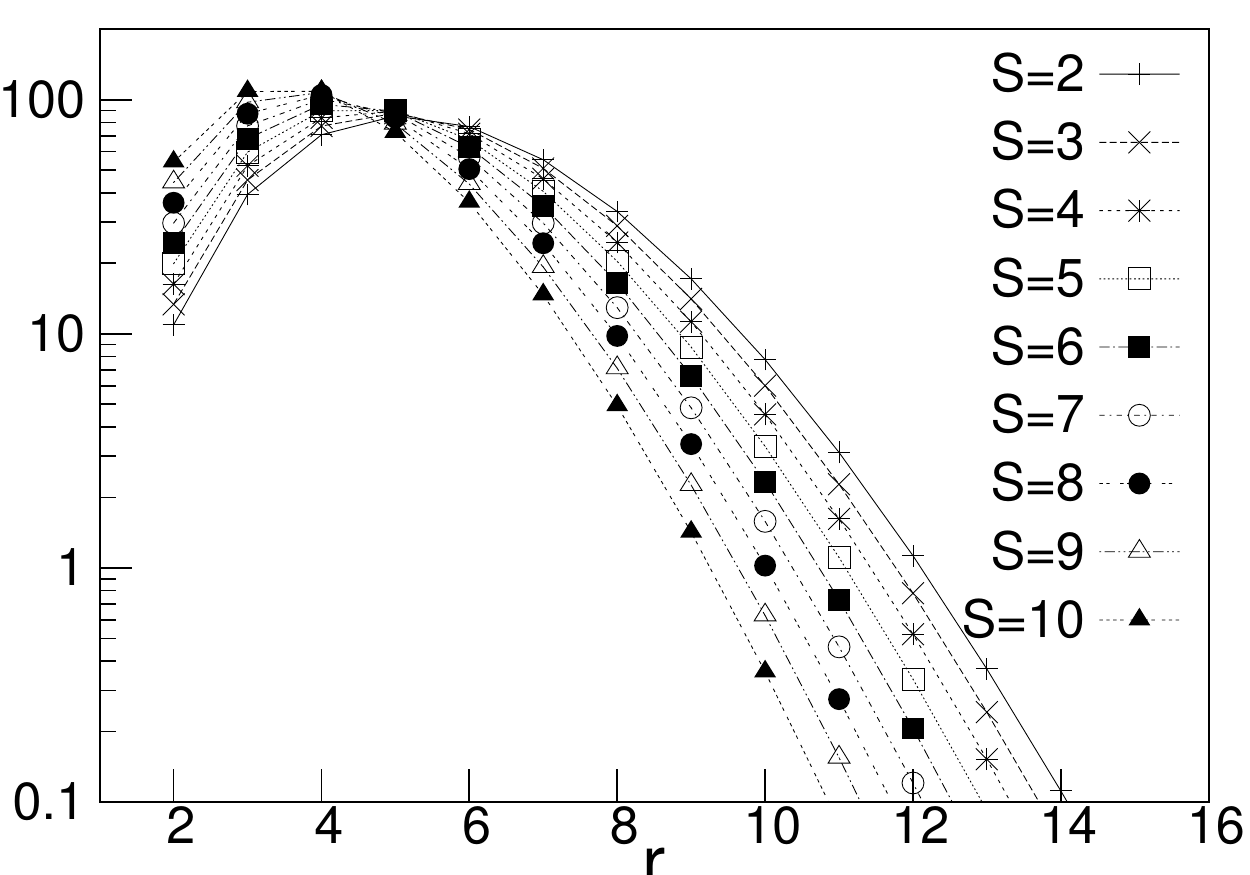}\\
  (a) & (b)\\
  \includegraphics[width=.47\textwidth]{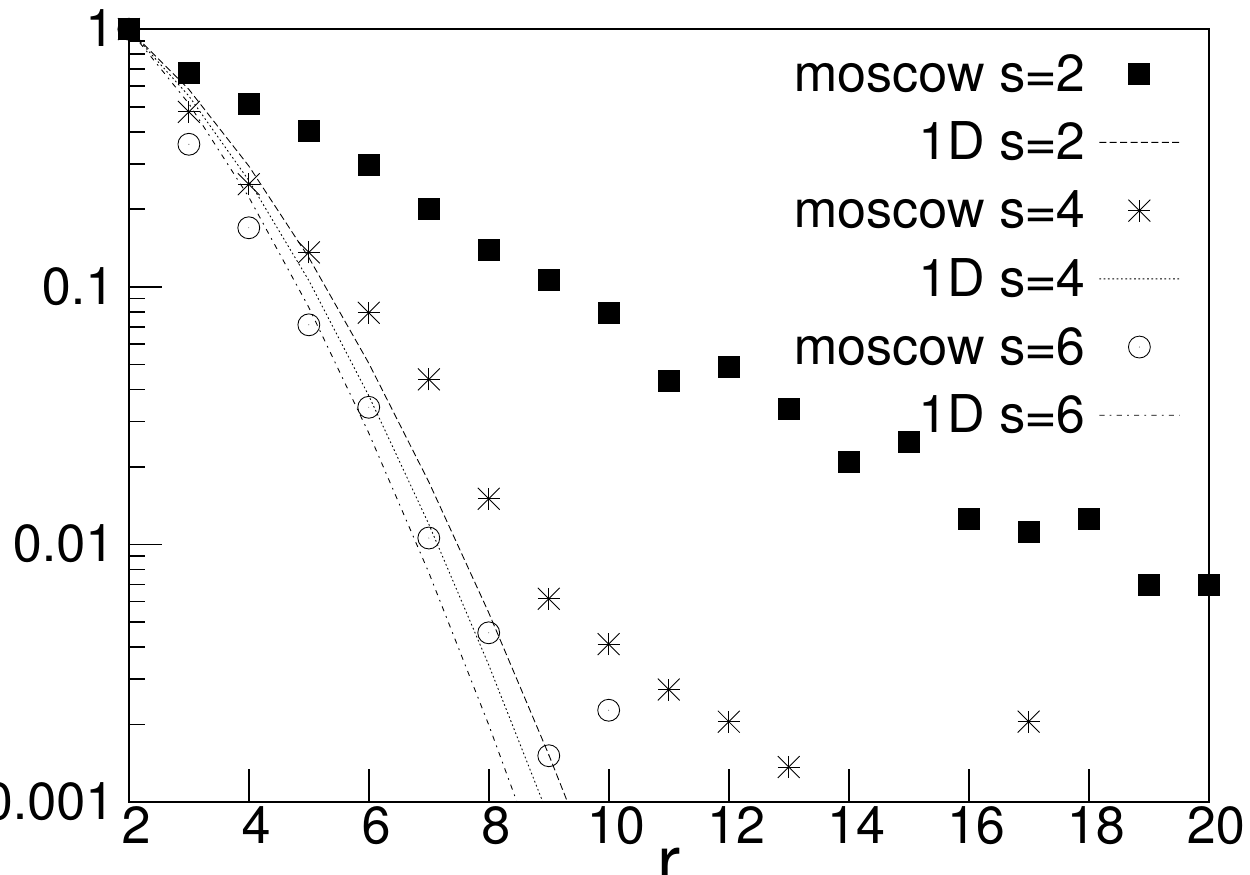} &
  \includegraphics[width=.47\textwidth]{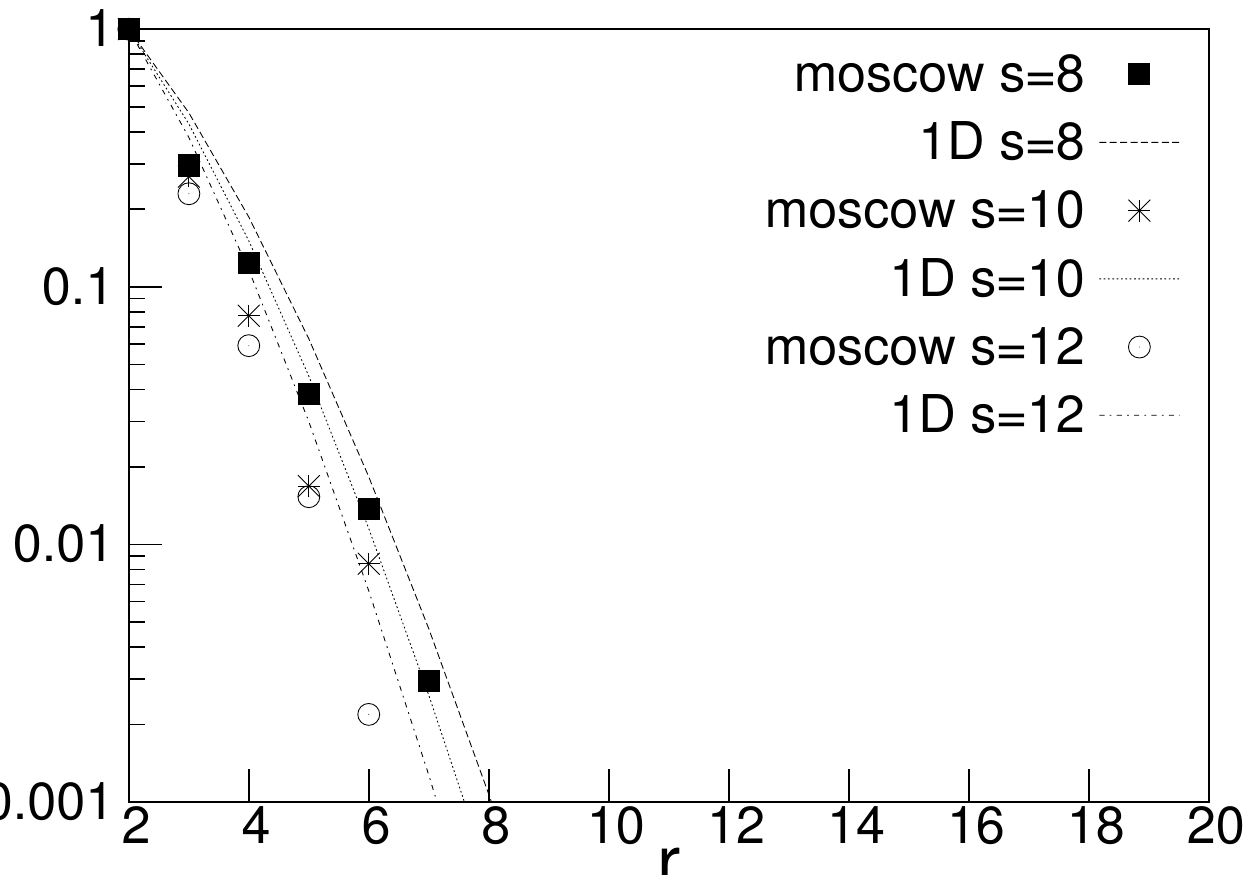}\\
  (c) & (d)
  \end{tabular}
  \caption{$s$-cumulative harness distribution $P_c(r,\hat s)$ as
 function of $r$ for fixed $\hat s$. Log-lin scale. a) Comparing
the analytical solution and numerical simulations for $N=10000$,
$R=1000$, S=$10$; b) for analytical solution for $N=10000$,
$R=2000$, S=$20$; c,d) comparing the analytical solution with
empirical results for the Moscow PTN ($R_{an}=R_{Moscow}=679$,
$S_{an}=\bar S_{Moscow}=22$, $N=5250$) for different $s$ normalized
by $P(2,s)$.}
\end{figure}
 It
turns out that the average results fit this formula with very good
accuracy even for small $N$, for example $N=10$ and of course for
any larger values of $N$ (Fig. 4a). For a large range of parameters
$R$, $S$, $N$ the behavior of $P(r,s)$ looks similar to what is shown in
Fig. 4b. For high overall density $\rho\cdot S$ we observe that curves for different $s$
intersect at small $r$ as for some real-world PTNs.
In one dimension, this effect has the following explanation: when the space
is overcrowded one will in general find more than two routes to overlap for
small sequences of stations.

This model has three parameters: the number of routes $R$,
  the route length $S$ and the number of sites $N$. As is obvious
  from eq. (10)  the harness distribution $P(r,s)$
for all $r$ and $s$ is close to a Poisson decay (5) for any set of
parameters. Therefore PTNs with an exponential behavior of the
harness distribution $P(r,s)$ may be compared with the results of
the one-dimensional approach. As an example we compare the
  normalized harness distribution for Moscow and the one-dimensional set
  of lines (Fig. 4c,d), where the number of routes $R$ and the route length $S$
  were chosen to match those of the Moscow PTN ($S$ set to the average route length).

  However, the quantitative results for all observed PTNs are several
  orders of magnitude higher than the result obtained with (9) in one
  dimension for the same $R$ and $S$ (for any $N$). Furthermore, PTNs that
  show a power law behavior (1) are even qualitatively different from the random 1D
approach.
Another difference is that the harness
  distribution curves for different $s$ are very similar in shape and both slope
and curvature vary much less than for the PTN harness distributions.

In the following we test a
  two-dimensional model with the simple simulations.

\section{Simple 2d modeling}

It is obvious that simply throwing at random lines parallel to the
axis' of a 2d square lattice with periodic boundary conditions will
lead to the original 1d problem: If the lattice has $X\times X$
sites one would get $2X$ independent one-dimensional systems.
However, it is not a priory clear what results one will find for
more general sets of walks on a 2d square lattice.
\begin{figure}[*h]
\begin{tabular}{cc}
 \includegraphics[width=.47\textwidth]{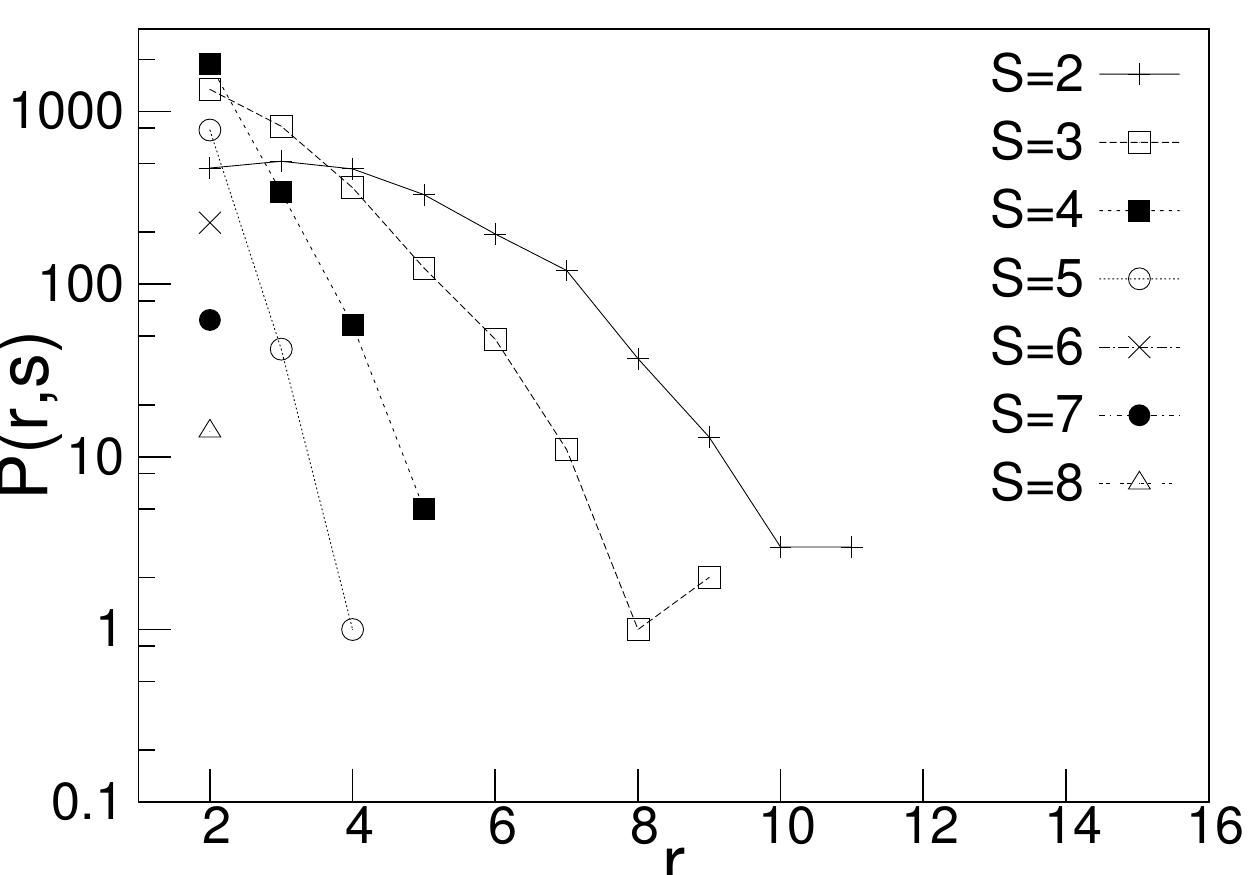} &
 \includegraphics[width=.47\textwidth]{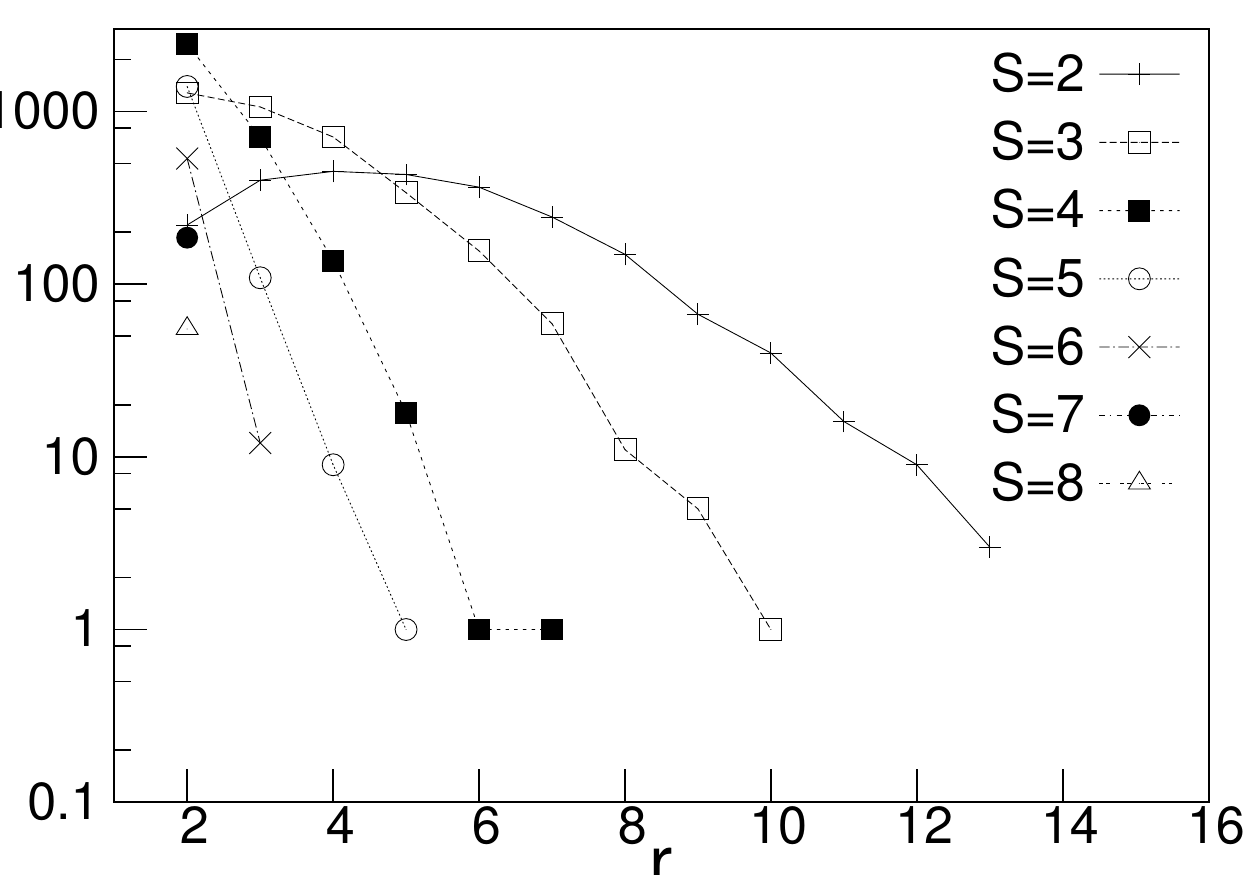}\\
 (a) & (b)\\
\includegraphics[width=.47\textwidth]{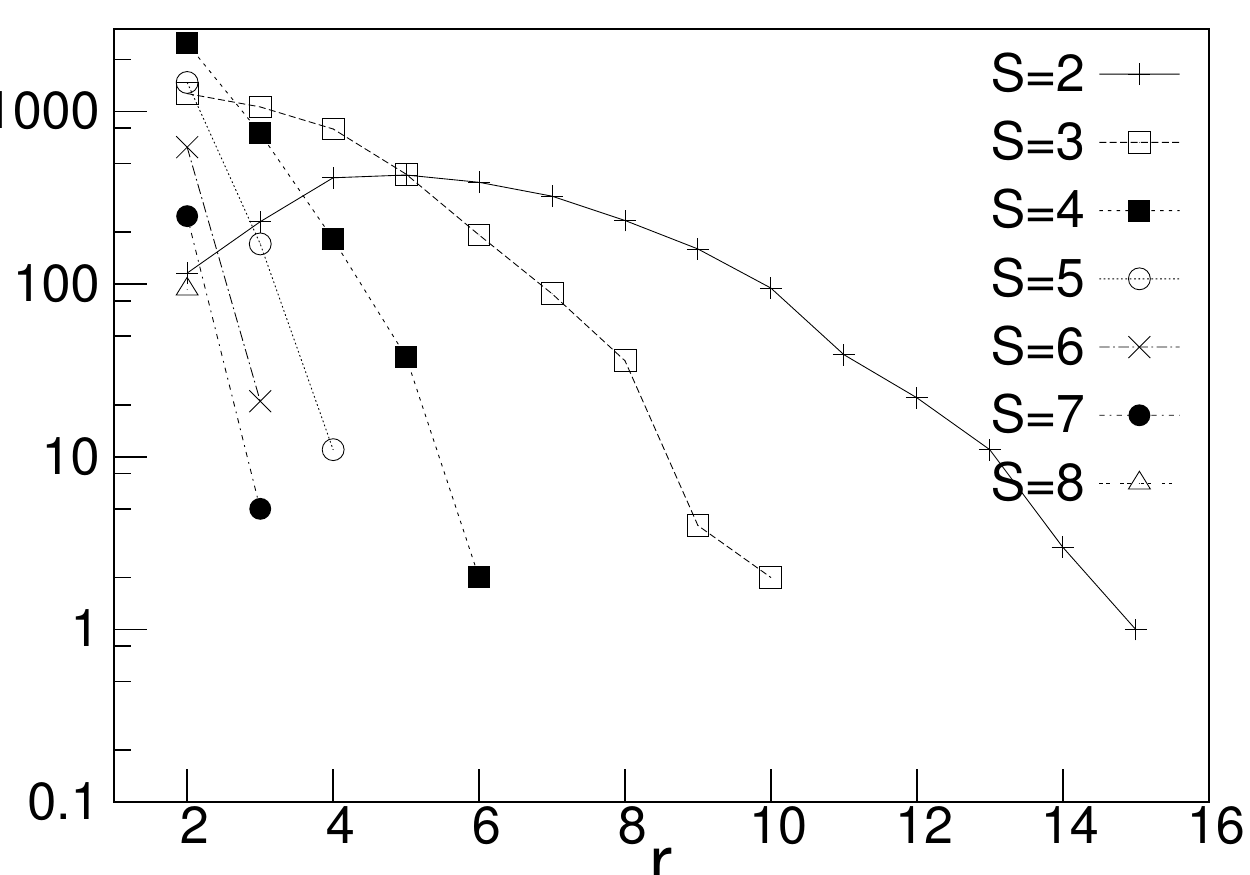} & (c)
 \end{tabular}
\caption{$s$-cumulative harness distribution $P_c(r,\hat s)$ as
function of $r$ for fixed $\hat s$ ($\hat s = 2\div 8$), for
$R=500$, $S=30$, $X=50$. Routes are generated as: a)RW  b)NRRW
c)SAW. $P(r,s)=0$ if $s>2$ for $r>11$ and if $s>7$ even for $r>2$.}
\end{figure}
To work this out, we implemented the following simulations.  We work
on a 2d $X\times X$ square lattice with periodic boundary
conditions. On this lattice we chose a set of $R$ walks each of
length $S$ (number of steps plus 1).  The routes are built either as
random walks (RW), non-reversal random walks (NRRW) that cannot
reverse the previous step, or self-avoiding random walks (SAW), that
may not intersect themselves.

These models have three parameters: the number of routes $R$, the
route length $S$ and the lattice size $X$. We choose the first two
parameters to match those of different real PTNs.

Postponing a more detailed analysis to a separate publication we
here summarize some of the main features of the harness distribution
$P_c(r,\hat s)$ of these models. Besides the finding that the
harness effect is "weak", some similarity between the the harness
effects seen in the three models is observed (Fig 5a,b,c). Curvature
and slope evolve in a similar way. Also intersections between the
curves for different $s$ are found to occur at lower values of $r$
in all cases. Differences are that the RW-generated networks
demonstrate a "weaker" harness effect, while NRRW- and SAW-generated
networks result in  harness distributions $P_c(r,\hat s)$ of similar
order of magnitude.

It turns out that for fixed $R$ and $S$, increasing the lattice size
$X$, the harness distributions $P(r,s)$, for all fixed $r<R$ and
$s<S$ show non-monotonous behavior (Fig. 6). As function of $X$ it
first increases, and then after reaching a maximum it starts to
decrease. Comparing with the empirical values found for real PTNs we
observe, that for some PTNs the empirical values even for small $r$
and $s$ are significantly larger than the maximum that could be
obtained with RW. For NRRW or SAW the empirical values are within
the observed range, however only for a small interval of $X$.

\begin{figure}[h]
  \includegraphics[width=.47\textwidth]{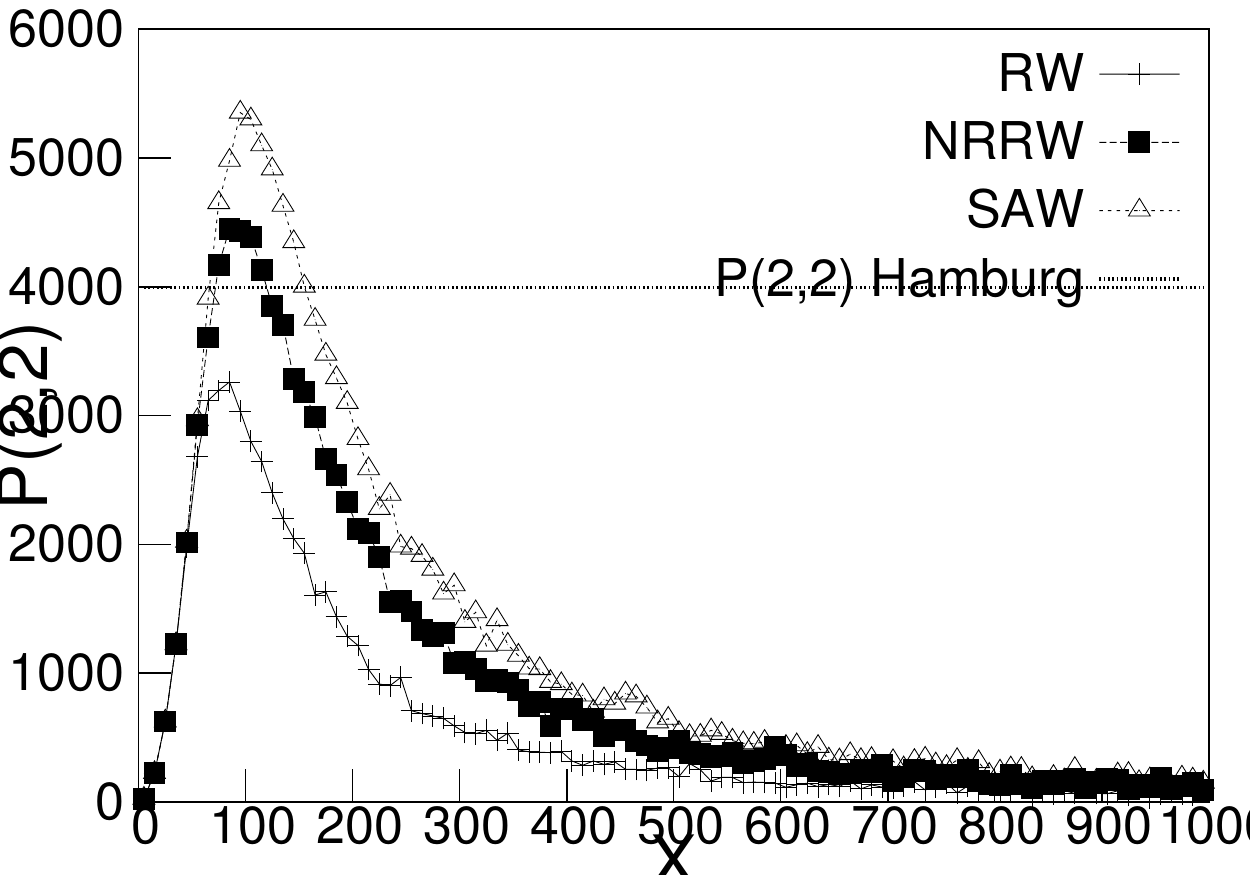}
  \caption{Situations observed for the $r$- and $s$-cumulative harness distribution
$P_{cc}(\hat r,\hat s)$ at $\hat r=2$, $\hat s=2$ as function of
lattice size $X$. For all three of models with $R=R_{Hamburg}$,
$S=\bar S_{Hamburg}$. The empirical value of $P_{cc}(2,2)$ for
Hamburg is shown for comparison.}
\end{figure}

This proves numerically the not surprising observation, that with
any of the proposed random or quasi-random walks only a very "weak"
harness effect may be obtained. In turn, this strongly indicates
that for most of the observed cities the harness effect must have a
structural background, that is not to be modeled by any of the
random approaches taken here.

Let us therefore analyze the harness effect for a model that
intrinsically takes correlations into account using some ideas of
non-equilibrium network evolution. [5].

\section{An evolutionary model in 2d}

From previous studies of PTNs [4, 5] the following facts are
relevant for our further discussion of PTN network evolution. The
final network often displays features (a power law degree
distribution) that are compatible with a preferential attachment
scenario [1]. Furthermore, the fractal dimension of public transport
routes appears to match that of 2d self-avoiding walks [5]. This
leads to the following model.

Given a 2d square lattice with side $X$ and periodic boundary
conditions, $R$ routes each of length $S$ will be created in the
following way. A first route is built as self-avoiding walk. Let
$k_{x}$ be the number of visits to lattice site $x$. The $R-1$
subsequent routes are constructed as SAWs with the following
preferential attachment rules:

a) choose a terminal station at $x_{0}$ with probability
\begin{equation}
  \label{f10}
  p \sim k_{x_0} + a/X^{2};
  \end{equation}

\begin{figure}[t]
  {a}\includegraphics[height=.15\textheight]{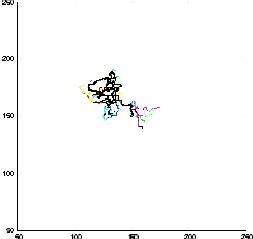}
  {b}\includegraphics[height=.15\textheight]{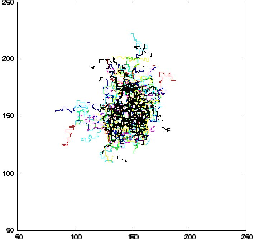}
  {c}\includegraphics[height=.15\textheight]{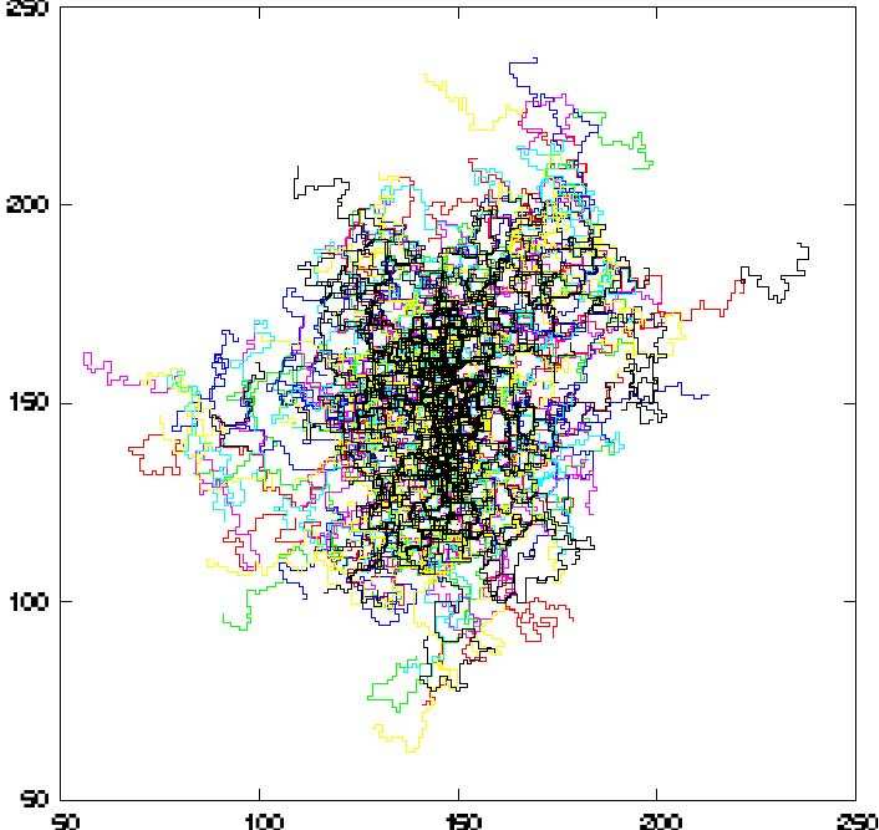}
  \caption{Simulated PTN maps. $X=300$, $R=1024$,
$S=64$. Parameter $a=0$. a) $b=0.1$  b) $b=0.3$  a) $b=0.5$. One can
see how parameter $b$ influences the spread of the network.}
\end{figure}
\begin{figure}[h]
  {a}\includegraphics[height=.15\textheight]{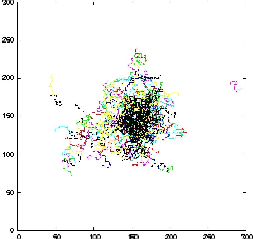}
  {b}\includegraphics[height=.15\textheight]{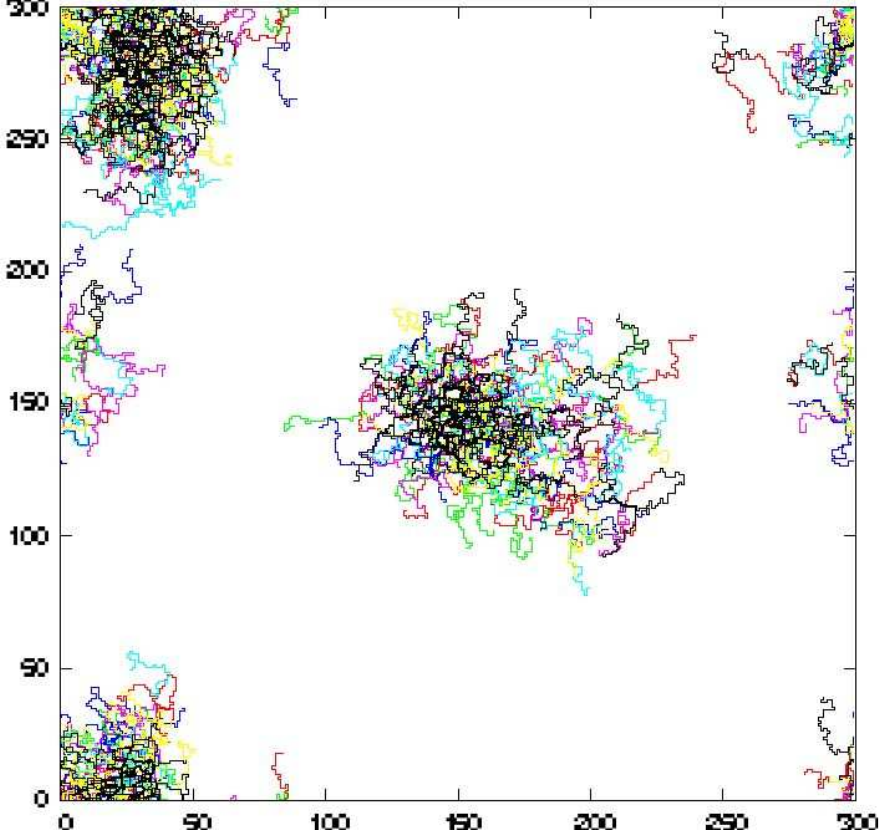}
  {c}\includegraphics[height=.15\textheight]{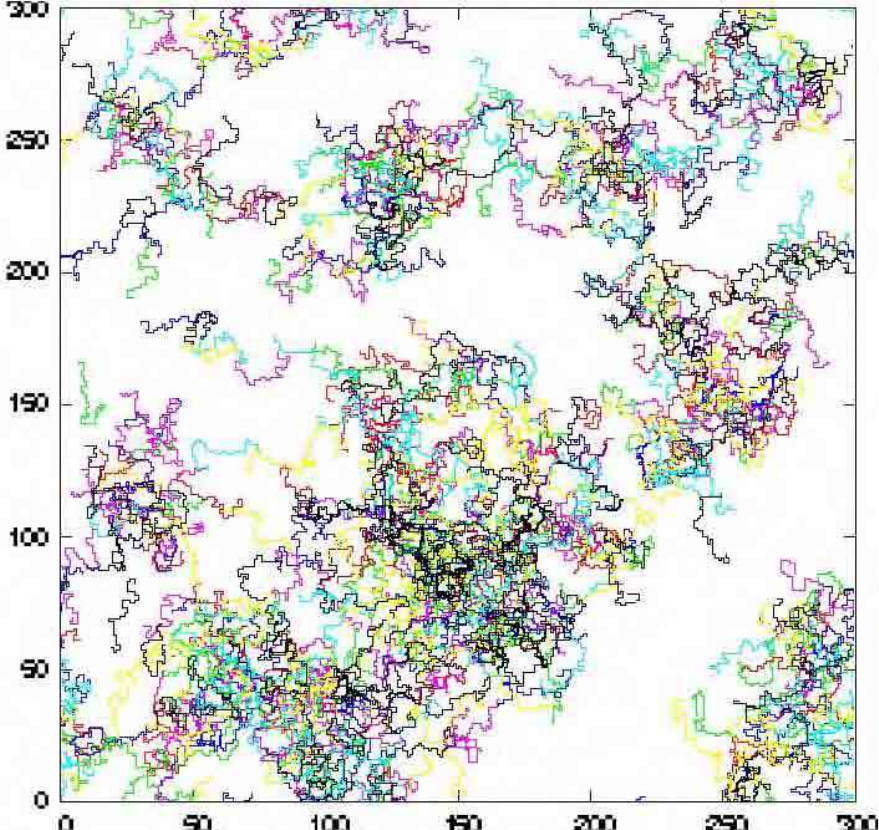}
  \caption{Simulated PTN maps. $X=300$, $R=1024$,
$S=64$. Parameter $b=0.5$. a) $a=15$  b) $a=20$  a) $a=500$. One can
see how parameter $a$ influences the number of clusters.}
\end{figure}

b) choose any subsequent station $x$ of the route with probability
\begin{equation}
  \label{f11}
  p \sim k_{x} + b.
  \end{equation}

c) If route intersects itself, discard it and return to step a).

Repeat steps a)-c) until $R$ routes are created.

We call this model the model of mutually interacting self-avoiding
walks (MI SAW). It implements a preferential attachment scenario for
networks that are built from chains (routes in our case). On the
figures presented (Fig. 7-8) one can see how the parameters $a$ and
$b$ influence the distribution of the routes on the lattice.
Parameter $a$ controls the number of disconnected clusters, while
parameter $b$ is related to the spread of each cluster. If both
parameters equal zero, all $R$ routes are restricted to the sites
occupied by the first route.

\begin{figure}[*h]
\begin{tabular}{cc}
  \includegraphics[width=.47\textwidth]{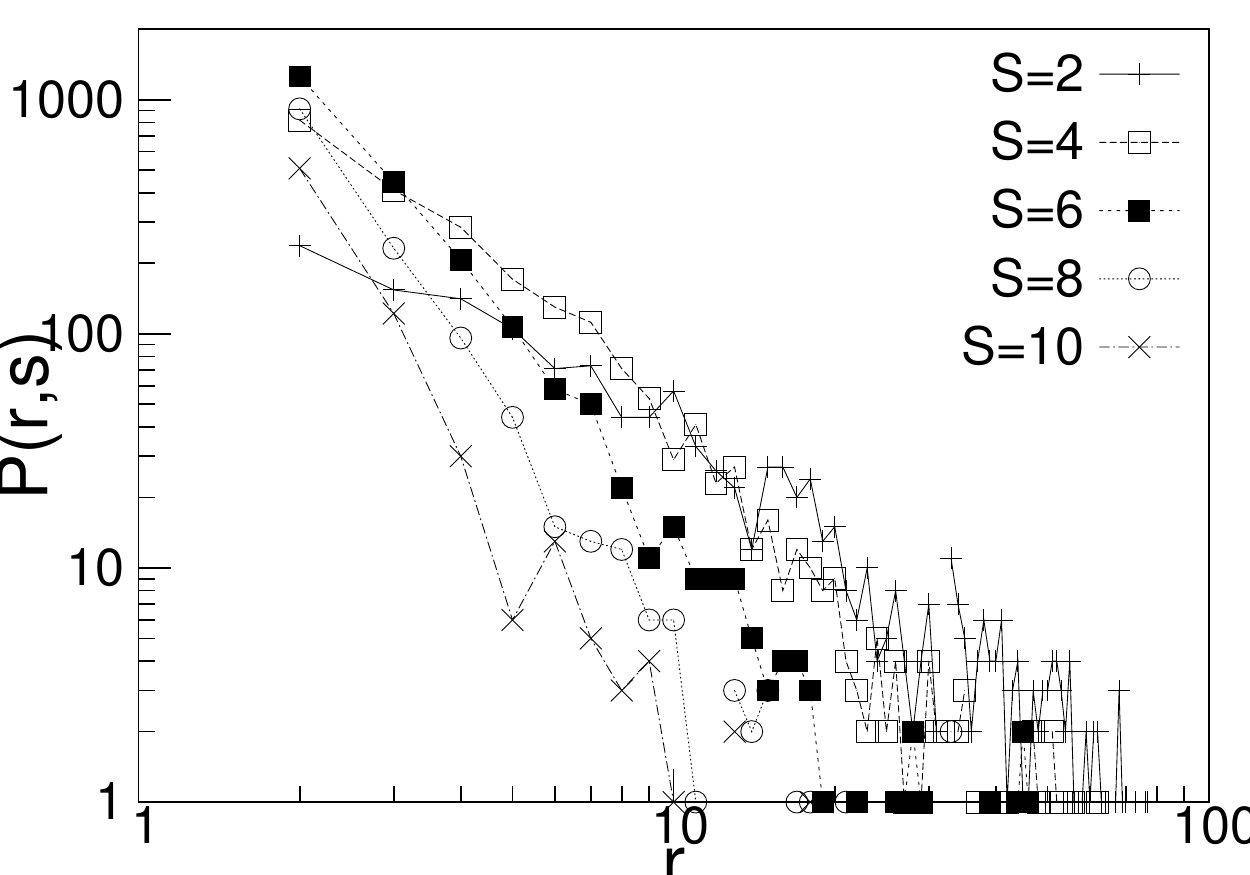} &
  \includegraphics[width=.47\textwidth]{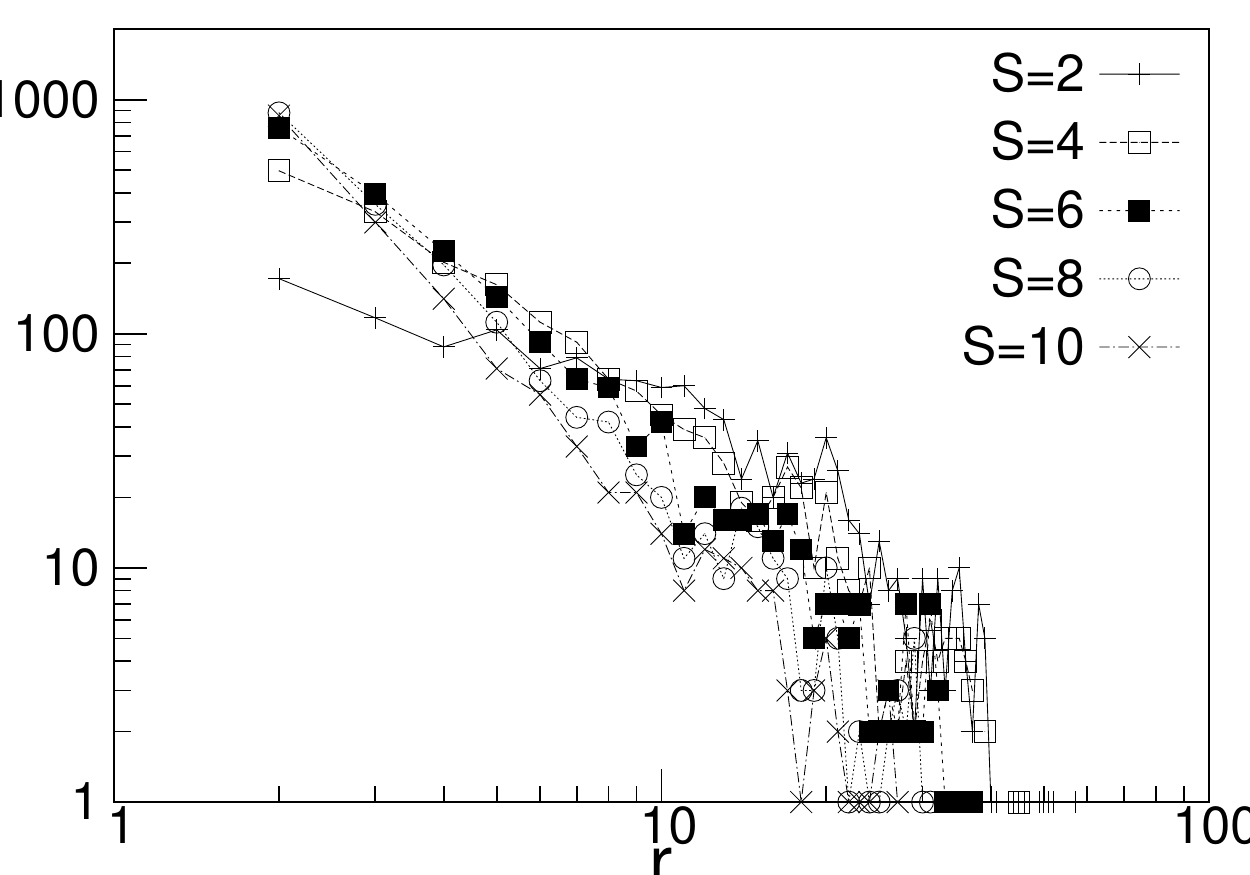} \\
  (a) & (b)\\
  \includegraphics[width=.47\textwidth]{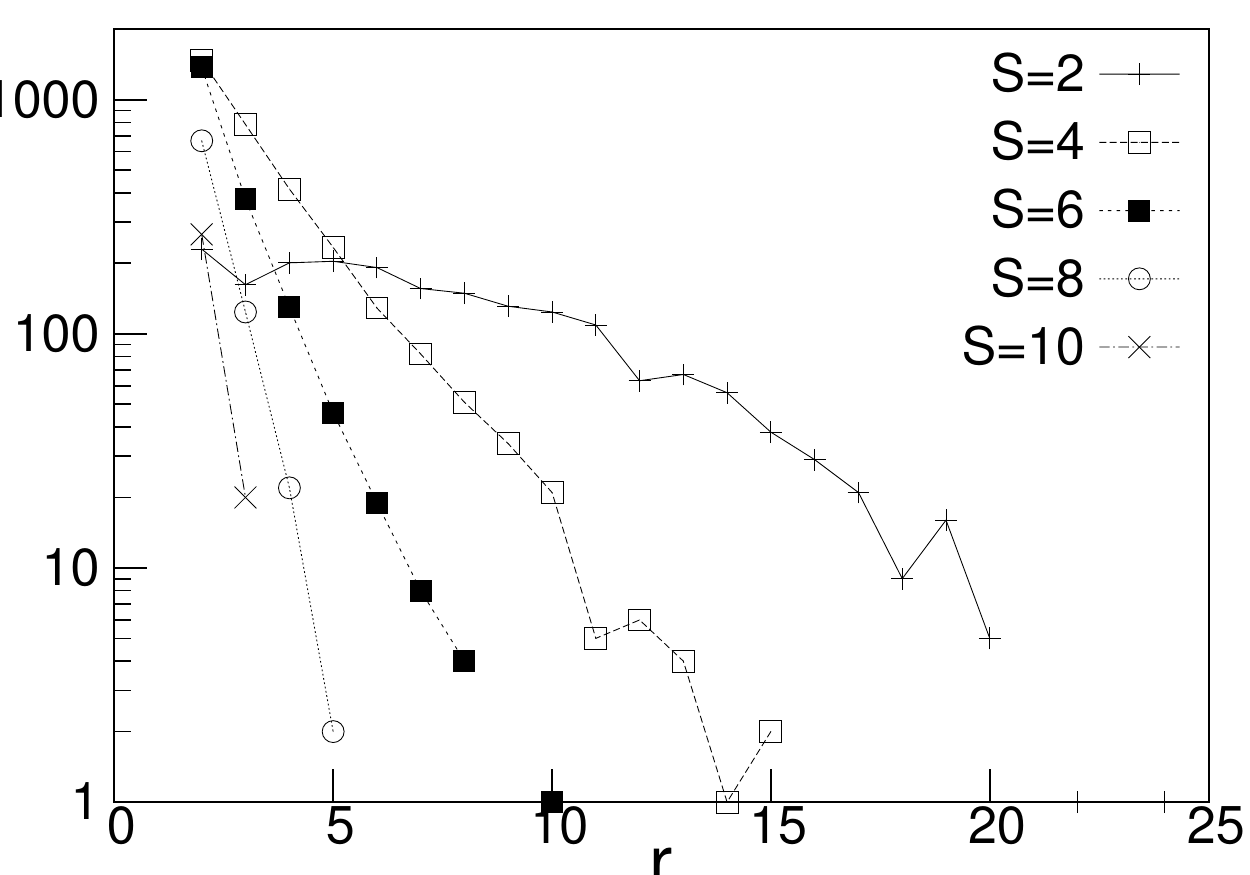} &
  \includegraphics[width=.47\textwidth]{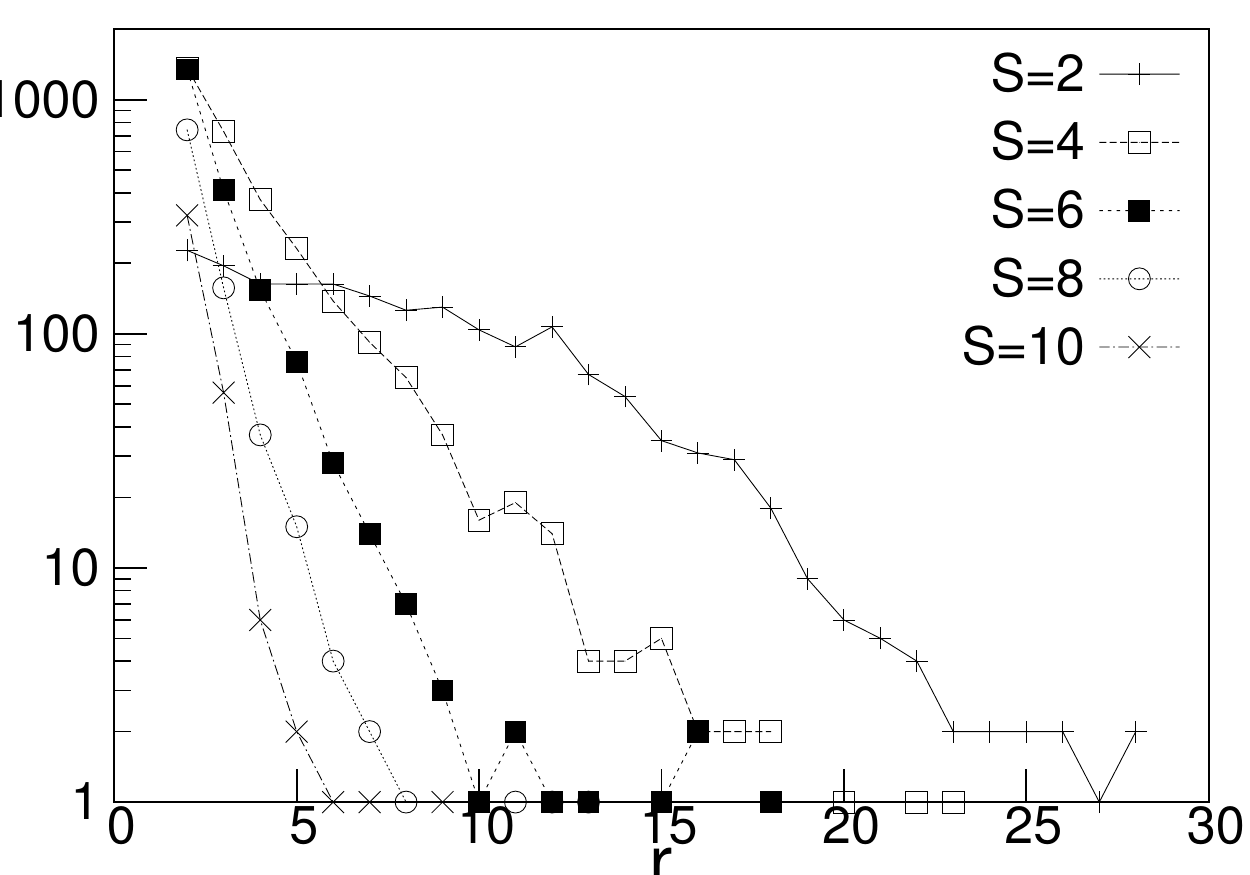} \\
  (c) & (d)
  \end{tabular}
  \caption{$s$-cumulative harness distribution $P_c(r,\hat s)$ as
function of $r$ for fixed $\hat s$ ($\hat s = 2,4,6,8,10$), for MI
SAW $R=500$, $S=30$, $X=50$. Log-log for a) a=0, b=0.5 and b) a=500,
b=0.1. Log-lin for c) a=15, b=0.5 and d) a=500, b=0.5.}
\end{figure}

To present the obtained results we show the harness distribution $P_c(r,\hat s)$
for different values for the model parameters $a$ and $b$ (Fig. 9). It
turns out that for choosing different values we may either find a behavior that
corresponds to a power law distribution (Fig. 9a,b) or a behavior that
corresponds to an exponential decay (Fig. 9c,d).

Also other features of real-world PTNs may be reproduced. With
increasing the sequence length $s$ the harness distribution $P(r,s)$
as a function of the number of routes $r$ attains a stronger
curvature and a steeper slope. Thus, with increasing sequence length
$s$ there is a cross-over from the power law regime (1) to an
exponential regime (2).

In the same way as observed for real-world PTNs in the 1d approach
and also for simple 2d models on the square lattice the
$s$-cumulative harness distributions $P_c(r,\hat s)$ as function of
the number of routes $r$ intersect at low values of $r$. In one
dimension we observed this behaviour because of an overcrowding
effect, see above. This also occurs to some extent in the 2d case,
however, on the 2d lattice there is an additional combinatorial
effect that leads to this result: the number of different possible
configurations of sequences with length $2$ is smaller than the
corresponding number of different possible configurations with
length $3$. To summarize one can say that the MI SAW model
reproduces a large amount of the empirically observed behavior of
harness distributions of PTNs.

\section{Conclusions}

Harness phenomena may be observed in different networks built
with space consuming links such as cables, vessels, pipes, neurons,
etc. The present analysis may possibly be applied also to such types of
networks. In the particular case of the PTNs that
we have analyzed we observe that in some cases the harness distribution may be
described by power laws. These observed power laws indicate a
certain level of organization and planning which may be driven by
the need to minimize the costs of infrastructure and secondly by the
fact that points of interest tend to be clustered in certain
locations of a city. This effect may be seen as a result of the
strong interdependence of the evolutions of both the city and its
PTN.

To further investigate the significance of the empirical results we
have studied one- and two-dimensional models of randomly placed
routes modeled by different types of walks. While in one dimension
an analytic treatment was successful, the two dimensional case was
studied by extensive simulations.

Our main results are the following:
\begin{itemize}
\item A one dimensional model for harness distributions
was solved analytically.
\item Exponentially decaying harness distributions may be reproduced by
the 1d approach.
\item Simple random placement of RW, SAW or NRRW on a two dimensional
square lattice result in weak harness distributions; in the RW case
much weaker than for real PTNs.
\item The $s$-cummulative distributions for different $s$ intersect
at low values of $r$ for all models due to combinatorial reasons.
\item A model of mutually interacting SAWs reproduces many of the
empirically observed features of harness distributions.
\end{itemize}

\begin{theacknowledgments}
We thank Yu. Holovatch for comments on the manuscript and on the
problem in general. T.H. is fully supported by Ecole Doctorale EMMA.
\end{theacknowledgments}



\end{document}